\documentstyle[seceq,twocolumn]{jpsj}
\newcommand{\rmd}{{\rm d}}
\newcommand{\iu}{{\rm i}}

\def\runtitle{Phase-Sensitive Impurity Effects}
\def\runauthor{Yusuke {\sc Kato}}

\title
{
Phase-Sensitive Impurity Effects in Vortex Core \\
of Moderately Clean Chiral Superconductors
}

\author
{ 
Yusuke {\sc Kato}\footnote{e-mail address: kato@coral.t.u-tokyo.ac.jp}
}
\inst
{
Department of Applied Physics, University of Tokyo, Tokyo 113-8656
}

\recdate{June 13, 2000}

\abst
{We study impurity effects in vortex core of two-dimensional moderately clean superconductors within the quasiclassical theory. The impurity scattering rate $\Gamma(E)$ of the Andreev bound states in vortex core with $+1$ vorticity of p-wave superconductors with ${\mib d}=\hat{\mib z}\left(p_x+\iu p_y\right)$ is suppressed, compared to the normal state scattering rate $\Gamma_{\rm n}$ in the energy region $\Gamma_{\rm n}^3/E_\delta^2\ll E\ll E_\delta\equiv |\delta_0|\Delta_\infty$ with scattering phase shift $\delta_0$ $(|\delta_0|\ll 1)$ and the pair-potential in bulk $\Delta_\infty$. 
Further we find that $\Gamma(E)/\Gamma_{\rm n}$ for p-wave superconductors with ${\mib d}=\hat{\mib z}\left(p_x-\iu p_y\right)$ is at most ${\cal O}(E/\Delta_\infty)$. These results are in marked contrast to the even-parity case (s,d-wave),  where $\Gamma(E)/\Gamma_{\rm n}$ is known to be proportional to $\ln(\Delta_\infty/E)$ . 
Parity- and chirality-dependences of impurity effects are attributed to the Andreev reflections involved in the impurity-induced scattering between bound states. Implications for the flux flow conductivity is also discussed. Novel enhancement of flux flow conductivity is expected to occur at $T\ll E_\delta$ for ${\mib d}=\hat{\mib z}\left(p_x+\iu p_y\right)$ and at $T\ll \Delta_\infty$ for  ${\mib d}=\hat{\mib z}\left(p_x-\iu p_y\right)$.
}
\kword
{Superconductivity, vortex, impurity effect, quasiclassical theory, flux flow conductivity}
\begin{document}
\sloppy
\maketitle
\section{Introduction}
Low energy physics in vortex state of superconductors without gap nodes is governed by localized excitations inside vortex cores.  The density of states within a core is comparable to that of the normal phase~\cite{Caroli} and hence vortex cores may be regarded as locally realized normal region. This picture has been used in the calculation of the flux flow conductivity in ref.~\citen{BardeenStephen}. Strictly speaking, however, the description of vortex core as a normal region is limited to the dirty superconductors, where the coherence length $\xi_0$ is much larger than the mean free path $l$. 

In clean superconductors where  $\xi_0\le l$ is satisfied, particles and holes within a vortex core are subject to Andreev reflections and the coherent superposition of particle and hole states form bound states before they are scattered by an impurity. The flux flow conductivity $\sigma_{\rm f}$ for clean s-wave superconductors has been calculated in refs.~\citen{BardeenSherman,LO76,KK} and found to be 
\begin{equation}
\sigma_{\rm f}/\sigma_{\rm n}\sim \ln\left(\Delta_\infty/T\right)H_{\rm c2}/B, 
\label{cleans}
\end{equation}  
in terms of the normal conductivity $\sigma_{\rm n}$, the upper critical field $H_{\rm c2}$ and magnetic induction $B$. The logarithmic factor in (\ref{cleans}) results from the shrinkage of vortex core at low temperatures (Kramer-Pesch effect)~\cite{KramerPesch} and logarithmic energy dependence of the impurity scattering rate of the Andreev bound states~\cite{KopninLopatin,KopninLopatin2}.        

Several clean superconductors are realized in correlated electron systems, such as ${\rm UPt_3}$, organic conductors, cuprates and ${\rm Sr_2 RuO}_4$. Those systems are believed to have unconventional superconducting states~\cite{review}. 

Flux flow conductivity of d-wave superconductors has been calculated in refs.~\citen{KopninVolovik, Mahklin} and it is similar to the s-wave result (\ref{cleans}) in moderately clean regime, where
\begin{equation}
l \Delta_\infty/\epsilon_{\rm F}\ll \xi_0\le l,\quad\left(\epsilon_{\rm F};\mbox{ Fermi energy}\right)
\label{moderatelyclean}
\end{equation}
is satisfied. Thus the flux flow conductivity of even-parity superconductors are expected to be similar.
  
However, flux flow conductivity of odd-parity superconductors may be different. Actually, it has been shown in ref.~\citen{Volovik} that single impurity does not change the spectrum within a vortex of a chiral superconductor when only the backward scattering is considered. Further, from the random-matrix approach, level statistics within a p-wave vortex core is expected to be different from that within an s-wave core~\cite{Bocquet,Ivanov}. Even experimentally, appreciable deviation from (\ref{cleans}) has been reported in the measurement in ${\rm UPt}_3$, where odd-parity superconducting states are presumably realized~\cite{Kambe,Lutke}.
Thus novel impurity effects in vortex core and flux flow conductivity are expected in odd-parity superconductors. 

For last five years, Kopnin {\it et al.}~\cite{KopninLopatin,KopninLopatin2}, Stone~\cite{Stone} and Blatter {\it et al.}~\cite{Blatter} developed the kinetic theory for vortex dynamics in clean superconductors. Their results on flux flow conductivity in the moderately clean regime reduce to  
\begin{equation}
\sigma_{\rm f}\sim n e\omega_0 \tau/B,  
\label{sigmaf2}
\end{equation}
with electron density $n$, charge unit $e(>0)$ and the relaxation time $\tau$ and angular velocity $\omega_0$ of precession of the Andreev bound states within vortex cores~\cite{Stone}. Since $\omega_0$ is expected to be of the order of $\ln\left(\Delta_\infty/T\right)\Delta_\infty^2/\epsilon_{\rm F}$, the calculation of $\sigma_{\rm f}$, essentially, reduces to that of relaxation time $\tau$; impurity problem inside vortex cores.  

Although several authors~\cite{Volovik,Bocquet,Ivanov} addressed the impurity problem within vortex of odd-parity superconductors, microscopic theory valid in the moderately clean regime (\ref{moderatelyclean}) is still missing~\cite{note}. Such a theory is highly desirable in the following reasons: first, an existing odd-parity superconductor ${\rm Sr_2RuO_4}$ has~\cite{Mackenzie} $\Delta_\infty/\epsilon_{\rm F}\sim 10^{-4}$ and $l/\xi_0\sim 8$ and thus belongs to the moderately clean regime. 
Another odd-parity one UPt$_3$ has~\cite{Huxley} $\Delta_\infty/\epsilon_{\rm F}\sim 0.05$ and $l/\xi_0\sim 50$ and lies between (2) and superclean regime $(l\gg \xi_0\epsilon_{\rm F}/\Delta_\infty)$. Study in the moderately clean regime, therefore, has relevance to experiments. Second, it is not so clear to what extent the approximation adopted in ref.~\citen{Volovik} is valid. Third, the validity of the random-matrix approach~\cite{Bocquet,Ivanov} can be examined only from microscopic theory. 

In the present paper, we consider the impurity problem within pancake vortex cores of two-dimensional s-wave and chiral p-wave superconductors with ${\mib d}=\hat{\mib z}\left(p_x\pm \iu p_y\right)$ within the quasiclassical theory of superconductivity~\cite{Eilenberger,LO68,Eliashberg,SereneRainer}. Although the method to calculate the impurity scattering rate has been given in refs.~\citen{KopninLopatin, Kopnin99}, we develop an alternative analytical method, which is based on Riccati formulation of the quasiclassical theory~\cite{Nagato1,Higashitani,Nagato2,SchopohlMaki,Schopohl,Eschrig,Eschrig2}.

In the next section, we present the framework of quasiclassical theory of superconductivity. In \S 3, impurity problem of s-wave vortex core is considered. In  \S 4, the results on p-wave case are given.  In \S 5, a brief argument on the flux flow conductivity is given. In \S 6, we discuss physical origin of the parity- and chirality- dependent impurity scattering rate inside vortex core. We also discuss the formulation used in \S 3 and \S 4. 
In \S 7, we summarize the present study.       
   
\section{Quasiclassical Theory of Superconductivity}
We consider two-dimensional superconductors with circular Fermi surface in type II limit. Quasiclassical theory of superconductivity~\cite{Eilenberger,LO68,Eliashberg,SereneRainer} is described in the equilibrium case by the quasiclassical green function
\begin{equation}
\hat g(z,\hat {\mib p},{\mib r})=\left(
\begin{array}{rc}
g&f\\
-\tilde f&-g
\end{array}\right),
\label{greenfunction}
\end{equation}
which is a $2\times 2$ matrix in particle-hole space and is a function of (complex) frequency $z$, direction $\hat{\mib p}=(\cos\alpha,\sin\alpha)$ of momentum ${\mib p}=p_{\rm F}\hat{\mib p}$, a point ${\mib r}=r(\cos \phi,\sin \phi)$ in real space. The equation of motion for $\hat g$ is given by
\begin{equation}
-\iu v(\hat{\mib p})\cdot{\mib\nabla} \hat g=\left[\epsilon \hat\tau_3-\hat \Delta({\mib r},\hat {\mib p})-\hat\Sigma,\hat g\right],
\label{qclequation}
\end{equation}
supplemented by the normalization condition 
\begin{equation}
\hat g^2=-\pi^2 \hat 1, 
\label{normalization}
\end{equation}
where $\hat 1$ is the $2\times 2$ unit matrix. 

Here $\hat \tau_3$ is a Pauli matrix,
\begin{equation}
\left(
\begin{array}{rc}
1&0\\
0&-1
\end{array}\right)
\label{pauli3}
\end{equation}
and $\hat \Delta$ is given by
\begin{equation}
\left(
\begin{array}{rc}
0&\Delta({\mib r},\hat{\mib p})\\
-\Delta^*({\mib r},\hat{\mib p})&0
\end{array}\right), 
\label{deltahat}
\end{equation}
where $\Delta({\mib r},\hat{\mib p})$ is the pair-potential. The impurity self-energy $\hat\Sigma$ in the t-matrix approximation is given by
\begin{equation}
\hat \Sigma=\hat \Sigma(z,{\mib r})=\frac{n_{\rm i}V}{1-N_0 V\langle\hat g\rangle}
, 
\label{tmatrix}
\end{equation}
with impurity concentration $n_{\rm i}$,  the density of states in the normal phase $N_0$ and the scattering potential for a single electron $V$.  
%
%
The symbol $\langle\cdots \rangle$ denotes the average of $\cdots$ over the Fermi surface. 

For later convenience, we parameterize the self-energy (\ref{tmatrix}) in terms of impurity scattering rate of normal phase
\begin{equation}
\Gamma_{\rm n}=\frac{1}{2\pi \tau_{\rm n}}=\frac{n_{\rm i}N_0 V^2}{1+\left(\pi N_0 V\right)^2}
\end{equation} 
and the scattering phase shift
$
\tan \delta_0 =-\pi N_0 V.$ Expression (\ref{tmatrix}) then becomes
\begin{equation}
\hat \Sigma\sim \frac{\Gamma_{\rm n}\langle\hat g\rangle}{\cos^2\delta_0 +\pi^{-2}\sin^2\delta_0\left(\langle f \rangle \langle \tilde f \rangle-\langle g \rangle^2\right)}, 
\label{Sigmatmatrix}
\end{equation}
where we have dropped the term proportional to $\hat 1$; it gives no contribution in (\ref{qclequation}). In the Born limit $(\delta_0\rightarrow 0)$, the self-energy reduces to 
\begin{equation}
\hat \Sigma=\Gamma_{\rm n}\langle\hat g\rangle.  
\label{Sigma}
\end{equation}

In this paper, we consider the case where $\hat g$ is an analytic function of $z$ in the upper half complex plane. Setting $z=\epsilon+\iu\delta$ with real $\epsilon$, we obtain the retarded Green function. 

In our case, the Fermi surface is circular and hence, the Fermi velocity ${\mib v}(\hat{\mib p})$ is given by $v\hat {\mib p}$ with a constant $v$. The left hand side of (\ref{qclequation}) then becomes $-\iu v \partial \hat g/\partial s$ with 
\begin{equation}
s={\mib r}\cdot\hat{\mib p}=r\cos\left(\phi-\alpha\right).
\label{s}
\end{equation} 
Thus, the direction of momentum specifies a line (^^ ^^ quasiclassical trajectory"~\cite{SereneRainer}), on which the quasiclassical equation is to be solved. Each quasiclassical trajectory can be determined by the condition $b({\mib r},\alpha)=$ constant, where the impact parameter $b({\mib r},\alpha)$ is defined as
\begin{equation}
b({\mib r},\alpha)={\mib r}\cdot \left(\hat{\mib z}\times\hat {\mib p}\right)=r\sin\left(\phi-\alpha\right).
\label{b}
\end{equation}

In this paper, we solve the quasiclassical equation (\ref{qclequation}) using a special parameterization of the quasiclassical Green function~\cite{Nagato1,Higashitani,Nagato2,SchopohlMaki,Schopohl,Eschrig}. The solution $\hat g$ of (\ref{qclequation}) can be written as
\begin{equation}
\hat g=\frac{-\iu\pi}{1+\gamma\gamma^\dagger}
\left(
\begin{array}{cc}
1-\gamma\gamma^\dagger&2\gamma\\
2\gamma^\dagger&-1+\gamma\gamma^\dagger
\end{array}\right).
\label{twobytwo}
\end{equation}
Here $\gamma$ and $\gamma^\dagger$ are the solutions of the following Riccati differential equations: 
\begin{subeqnarray}
&\iu {\mib v}\cdot{\mib \nabla}\gamma&=-\left(\Delta^*-\Sigma_{21}\right)\gamma^2-2\left(\epsilon-\Sigma_{11}\right)\gamma-\left(\Delta+\Sigma_{12}\right)
\label{Riccati1}\\
&\iu {\mib v}\cdot{\mib \nabla}\gamma^\dagger&=-\left(\Delta+\Sigma_{12}\right)\gamma^{\dagger 2}+2\left(\epsilon-\Sigma_{11}\right)\gamma^\dagger-\left(\Delta^*-\Sigma_{21}\right). 
\label{Riccati2}
\end{subeqnarray}
In this parameterization, the normalization condition (\ref{normalization}) is automatically satisfied. 
Now we fix $\epsilon$, $\alpha$ and $b=r\sin(\phi-\alpha)$. Along the quasiclassical trajectory, equations~(\ref{Riccati1}a,b) are to be solved respectively, under the initial conditions:
 
\begin{subeqnarray}
&\gamma&=\frac{\Delta}{-\epsilon-\iu \delta-\iu\sqrt{\left|\Delta\right|^2-\epsilon^2}}, \quad\mbox{ for }s=-s_{\rm c}
\label{ic1}\\
&\gamma^\dagger&=\frac{\Delta^*}{\epsilon+\iu \delta+\iu\sqrt{\left|\Delta\right|^2-\epsilon^2}}, \quad\mbox{ for }s=s_{\rm c},
\label{ic2}
\end{subeqnarray}
where $s_{\rm c}$ is a positive number much larger than $\xi_{0}$ and eventually the limit $s_{\rm c}\rightarrow +\infty$ will be taken. 
The initial conditions~(\ref{ic1}a,b) are obtained from the bulk solution of (\ref{qclequation}). In those initial conditions, the effect of nonmagnetic impurity is absent as a result of Abrikosov-Gorkov theory for isotropic s-wave homogeneous superconductors~\cite{AG}.
  
We consider a single vortex with $+1$ vorticity using the following test pair-potentials:
\begin{eqnarray}
& &\Delta({\mib r},\hat{\mib p})\nonumber\\
&=&\Delta_0(r) {\rm e}^{\iu \phi}\left\{
\begin{array}{ll}
1& \mbox{for s-wave superconductors(SC)}\\
{\rm e}^{\pm \iu\alpha}& \mbox{for p-wave SC } ({\mib d}=\hat{\mib z}(p_x\pm\iu p_y)).
\end{array}
\right.
\end{eqnarray} 
The amplitude of the pair-potential $\Delta_0(r)$ is assumed to be a function of $r=|{\mib r}|$ (axisymmetric) and behave as
\begin{equation}
\Delta_0(r)=\left\{
\begin{array}{cc}
\Delta_\infty r/\xi_1,&\mbox{ for }r\ll \xi_1\\
\Delta_\infty ,&\mbox{ for }r\gg \xi_1\\
\end{array}
\right.
\label{xi1}
\end{equation}
with a characteristic length $\xi_1$ and the amplitude of the pair-potential in the bulk $\Delta_\infty$. We take $\xi_1 (\le \xi_0)$ as a length different from the conventional coherence length $\xi_0=v/(\pi \Delta_\infty)$ in general, in order to take account of the shrinkage of a vortex core~\cite{KramerPesch}. 
\section{Isotropic S-wave Case}
\subsection{Self-energy in Born limit}
In pure superconductors ($\hat \Sigma=0$), the approximate expression for $\hat g$ has been obtained~\cite{KramerPesch, Eschrig}. For $\epsilon\ll \Delta_\infty$ and $b\ll \xi_0$, 
\begin{equation}
\hat g=\frac{\hat g_0(\epsilon,\alpha,s,b)}{\epsilon +\iu \delta -E(b)}+\mbox{(regular part)},
\label{gapproximate}
\end{equation}
Here the spectral function $\hat g_0(\epsilon,\alpha,s,b)$ has no singularity and varies with energy over the range of $\Delta_\infty$ and with s over that of $\xi_0$. $E(b)$ is the pole corresponding to the Andreev bound state for a given quasiclassical trajectory. 
\begin{equation}
E_{\rm p}(b)=\frac{b}{C}\int_0^\infty\rmd s\frac{\Delta_0(s)}{s}{\rm e}^{-u(s)}
\sim \frac{2b\Delta_\infty^2}{v}\ln\left(\frac{\pi\xi_0}{2\xi_1}\right)\label{Epb}
\end{equation}
with
\begin{equation}
u(s)=\frac{2}{v}\int_0^{\left|s\right|}\rmd s' \Delta_0(s') 
\label{u}
\end{equation}
and the normalization
\begin{equation}
C=\int_0^\infty\rmd s\exp\left[-u(s)\right]\sim \pi\xi_0/2  
\end{equation}
(the subscript ^^ ^^ p" in ({\ref{Epb}}) stands for pure superconductors). 
The second term in the right hand side of (\ref{gapproximate}) comes from the scattering state of quasiparticle. This contribution can be neglected in the case of $\epsilon \ll \Delta_\infty$. 

Next we consider the case with impurity. For ${\rm Max}\left(\epsilon,\Gamma\right) \ll\Delta_\infty$, main contribution to $\hat g$ comes from the Andreev bound states and an expression similar to (\ref{gapproximate}) still holds. In this case, however, those bound states are subject to the impurity scattering and the bound state pole moves to lower complex plane. Now we assume that $\hat g$ for
${\rm Max}\left(\epsilon,\Gamma_{\rm n}\right)\ll \Delta_\infty$ and $b\ll \xi_0$ has a form of 
\begin{equation}
\hat g\sim \frac{\hat g_0(\epsilon,\alpha,s,b;\Gamma_{\rm n})}{\epsilon +\iu \Gamma(b) -E(b)},
\label{gapproximate2}
\end{equation}      
with the weight function $\hat g_0$ varies with $\epsilon$ or $\Gamma_{\rm n}$ over the range of $\Delta_\infty$ and with $b$ over the range of $\xi_0$. The impurity scattering rate $\Gamma(b)$ for Andreev bound states appears in (\ref{gapproximate2}) as the imaginary part of pole of Green function.
 
Now we calculate the self-energy $\Sigma$ as a function of $(\epsilon,\alpha,{\mib r};\Gamma_{\rm n})$ by taking the average of (\ref{gapproximate2}) with respect to $\alpha$ for fixed ${\mib r}$. $s$, $b$ depend on $\alpha$ through (\ref{s}) and (\ref{b}), respectively. The weight function $\hat g_0$ can be replaced by the value for $\epsilon=b=\Gamma_{\rm n}=0$ with the accuracy of ${\cal O}(\epsilon/\Delta_\infty, \Gamma_{\rm n}/\Delta_\infty,b/\xi_0)$. This value is available from earlier results on pure case~\cite{KramerPesch,KopninLopatin,Eschrig}. The self-energy can then be written as
\begin{equation}
\hat\Sigma({\mib r};\Gamma_{\rm n})\sim \frac{\pi v\Gamma_{\rm n}}{2C}\langle\frac{\exp\left[-u(s)\right]\hat M_0}{\left(\epsilon-E(b)+\iu \Gamma(b)\right)}
\rangle,
\label{Sigmalowenergy}
\end{equation} 
 with (\ref{s}), (\ref{b}) and
\begin{equation}
\hat M_0=\hat M_0(\alpha)\equiv \left(
\begin{array}{ccc}
1&-\iu {\rm e}^{\iu \alpha}\\
-\iu {\rm e}^{-\iu \alpha}&-1
\end{array}\right). 
\label{M0}
\end{equation}
 
From now on, we assume that
\begin{equation}
E(b)\sim E_0+E' b,\quad 
E'v/\Delta_\infty\equiv\tilde\Delta_\infty\gg{\rm Max}\left(|\epsilon|,\Gamma\right).
\label{assumption1}
\end{equation}
with $b$-independent $E_0$ and $E'$ 
and
\begin{equation}
\Gamma(b)\ll E'. 
\label{Gammaassumption}
\end{equation}

The denominator in (\ref{Sigmalowenergy}) leads to a distribution localized around $\alpha=\alpha_1,\alpha_2$ satisfying 
\begin{equation}
\epsilon=E(b=r\sin(\phi-\alpha_i))\quad i=1,2.\label{s1s2}
\end{equation}
Therefore ^^ ^^ the slowly varying part "$\exp\left[-u(s)\right]$ can be replaced by 
$$
\exp\left[-u(s_{1,2}=r\cos(\phi-\alpha_{1,2}))\right].
$$ 
If we set $b_i\equiv r\sin(\phi-\alpha)$, then, (\ref{s1s2}) leads to $b_1=b_2$ and $s_1=-s_2$. It follows that $u(s_1)=u(s_2)$ because $u(s)$ is an even function of $s$ by definition (\ref{u}). The solutions (\ref{s1s2}), however, exist only if $r>\left({\rm Re}\epsilon-E_0\right)/E'$.
As a result, the self-energy (\ref{Sigmalowenergy}) becomes
\begin{equation}
\hat\Sigma(\epsilon, {\mib r};\Gamma_{\rm n})\sim \frac{\pi v \Gamma_{\rm n}\exp\left[-u(\tilde s_1)\right]}{2C}
\langle
\frac{\hat M_0\left(\alpha\right)}
{\tilde \epsilon-E' b}
\rangle,
\label{Sigma2}
\end{equation} 
with $\tilde \epsilon\equiv \epsilon-E_0+\iu \Gamma$ and 
$$
\tilde s_1={\rm Max}\left[s_1,\left({\rm Re}\epsilon-E_0\right)/E'\right].
$$ 
 
A direct calculation shows that
\begin{equation}
\langle 1/(\tilde\epsilon -E'b)\rangle=
\frac{-i}{E'r\sqrt{1-w^2}}, 
\label{average1}
\end{equation}
with $w=\tilde\epsilon/(E' r)$.
Here the branch of $\sqrt{\cdots}$ is taken such that
\begin{equation}
\sqrt{1-w^2}\rightarrow
\left\{
\begin{array}{cc}
-\iu w,&\quad\mbox{ for }
\left|w\right|\gg 1\\
1,&\quad\mbox{ for }\left|w\right|\ll 1
\end{array}
\right.,
\label{branch}
\end{equation}

Similarly, we obtain 
\begin{equation}
\langle {\rm e}^{\pm\iu\alpha}/(\epsilon -E'b+\iu \Gamma)\rangle=
\frac{\mp\iu {\rm e}^{\pm \iu \phi}}{E'r\sqrt{1-w^2}\left(\iu w-\sqrt{1-w^2}\right)}
\label{average2}
\end{equation}
with the branch (\ref{branch}). 

By substituting (\ref{Sigma2}) with (\ref{average1}) and (\ref{average2}) into (\ref{qclequation}) and solving the resulting equation, we can obtain $E_0$, $E'$ and $\Gamma$. 
\subsection{Bound state pole in Born limit} 
Now we solve quasiclassical equation via (\ref{Riccati1}a,b) with a particular interest of the bound state pole. From (\ref{Riccati1}a,b), the equation $$
-\iu v\frac{\partial }{\partial s}\left(1+\gamma\gamma^\dagger\right)=\left\{\left(\Delta^*-\Sigma_{21}\right)\gamma+\left(\Delta+\Sigma_{12}\right)\gamma^\dagger\right\}\left(1+\gamma\gamma^\dagger\right)
$$
follows. This equation shows that the denominator of $\hat g$ vanishes for all $s$ on the trajectory if it vanishes at a point~\cite{Schopohl}. Thus the energy $\epsilon$ which gives $\left(1+\gamma\gamma^\dagger\right)=0$ has the meaning of a bound state pole even in the presence of impurity.  
 
What we need are the solutions for (\ref{Riccati1}a,b) in case of $\epsilon\ll \Delta_\infty$, $b\ll\xi_0$ and $\hat\Sigma\ll \Delta_\infty$.
The last inequality is equivalent to 
\begin{equation}
\Gamma_{\rm n}/\left|\tilde\epsilon\right|\ll 1
\label{assumption2}
\end{equation}
because 
$$
\Sigma\sim \frac{\pi\Gamma_{\rm n}}{2C\left|\tilde\epsilon\right|},\quad C\sim \xi_0=\frac{v}{\pi\Delta_\infty}.
$$    

We expand equations (\ref{Riccati1}a,b) with $\epsilon$, $b$ and $\hat\Sigma$ up to first order.  It is convenient to introduce the following notations:
\begin{equation}
\begin{array}{ccc}
\gamma=\bar\gamma {\rm e}^{\iu \alpha},&\gamma^\dagger=\bar\gamma^\dagger {\rm e}^{-\iu \alpha},&\Delta=\bar\Delta {\rm e}^{\iu \alpha},\\
\Delta^*=\bar\Delta^\dagger {\rm e}^{-\iu \alpha},&\Sigma_{12}=\bar\Sigma_{12}{\rm e}^{\iu \alpha}, &\Sigma_{21}=\bar\Sigma_{21}{\rm e}^{-\iu \alpha}. 
\end{array}
\label{bargamma}
\end{equation}
First we consider the case where $\epsilon=0$, $b=0$ and $\bar\Sigma=0$. In this case, the solutions for (\ref{Riccati1}a,b) with initial conditions $\bar\gamma(s=-s_{\rm c})=-\iu$ and $\bar\gamma^\dagger(s=s_{\rm c})=-\iu$ are given by
\begin{equation}
\bar\gamma=\bar\gamma^\dagger=-\iu.
\end{equation}  
Next we consider the first order corrections with respect to  $\epsilon$, $b$ and $\bar\Sigma$. $\bar\gamma$ and $\bar\gamma^\dagger$ are expanded as
\begin{eqnarray}
\bar\gamma^{\left(\mbox{ },\dagger\right)}&=&-\iu+\epsilon\left(\partial_\epsilon\bar\gamma^{\left(\mbox{ },\dagger\right)}\right)+b\left(\partial_b\bar\gamma^{\left(\mbox{ },\dagger\right)}\right)+\bar\gamma^{\left(\mbox{ },\dagger\right)}_\Sigma\nonumber\\
&+&{\cal O}\left(\epsilon^2,b^2,\Sigma^2,\epsilon b,\epsilon\Sigma,b\Sigma\right),   
\label{expand}
\end{eqnarray}
where
\begin{equation}
\partial_\epsilon\bar\gamma^{\left(\mbox{ },\dagger\right)}=\left.\frac{\partial \bar\gamma^{\left(\mbox{ },\dagger\right)}}{\partial \epsilon}\right|_{\epsilon=b=\Sigma=0},\quad \partial_b\bar\gamma^{\left(\mbox{ },\dagger\right)}=\left.\frac{\partial \bar\gamma^{\left(\mbox{ },\dagger\right)}}{\partial b}\right|_{\epsilon=b=\Sigma=0}.
\label{partialgamma}
\end{equation}
Here $\bar\gamma^{\left(\mbox{ },\dagger\right)}$ is either $\bar\gamma$ or $\bar\gamma^\dagger$. 
The symbol $\bar\gamma^{\left(\mbox{ },\dagger\right)}_\Sigma$ denotes the solution linear order of $\Sigma$ in the following situation: $b$ is taken to be zero. $\epsilon$ as a coefficient of $\bar\gamma^{\left(\mbox{ },\dagger\right)}$ is set to be zero while the implicit energy dependence of $\Sigma$ is kept.   
The quantities (\ref{partialgamma}) are given by~\cite{Eschrig}
\begin{equation}
\partial_\epsilon\bar\gamma^{\left(\mbox{ },\dagger\right)}={\rm e}^{u(s)}\left[\Delta_\infty^{-1} {\rm e}^{-u(s_{\rm c})}\pm 2v^{-1}\int^s_{\mp s_{\rm c}}\rmd s'{\rm e}^{-u(s')}\right]
\label{gammaepsilon}
\end{equation}
and
\begin{equation}
\partial_b\bar\gamma^{\left(\mbox{ },\dagger\right)}={\rm e}^{u(s)}\left[-s_{\rm c}^{-1} {\rm e}^{-u(s_{\rm c})}\mp 2v^{-1}\int^s_{\mp s_{\rm c}}\rmd s' s'^{-1}\bar \Delta(s'){\rm e}^{-u(s')}\right]. \label{gammab}
\end{equation}
In (\ref{gammaepsilon}) and (\ref{gammab}), the upper signs are respectively for $\partial_{\epsilon}\bar\gamma$ or $\partial_{b}\bar\gamma$ and the lower one for $\partial_{\epsilon}\bar\gamma^\dagger$ or $\partial_{b}\bar\gamma^\dagger$.

Now we consider $\bar\gamma_{\Sigma}^{(\mbox{ },\dagger)}$. They are the solutions of
\begin{equation}
v\frac{\partial \bar\gamma_{\Sigma}}{\partial s}=2\bar\Delta\bar\gamma_\Sigma-2\tilde\Sigma
,\quad v\frac{\partial \bar\gamma^{\dagger}_\Sigma}{\partial s}=2\bar\Delta\bar\gamma_\Sigma^{
\dagger}+2\tilde\Sigma,
\label{gammasigmaequation}
\end{equation} 
with 
\begin{equation}
\tilde\Sigma\equiv \Sigma_{11}-\iu\left(\bar\Sigma_{12}+\bar\Sigma_{21}\right)/2. 
\label{tildesigma}
\end{equation}
Equations (\ref{gammasigmaequation}) are supplemented by the initial conditions 
\begin{equation}
\bar\gamma_\Sigma(s=-s_{\rm c})=0,\quad \bar\gamma_\Sigma^\dagger(s=s_{\rm c})=0.
\label{icgammasigma}
\end{equation}
Equations (\ref{gammasigmaequation}) with (\ref{icgammasigma}) yield the following solutions: 
\begin{equation}
\bar\gamma^{(\mbox{ },\dagger)}_\Sigma=\mp 2v^{-1}{\rm e}^{u(s)}\int_{\mp s_{\rm c}}^s\rmd s'{\rm e}^{-u(s')}\tilde\Sigma(s').
\label{gammasigma}
\end{equation}
After taking the limit $s_{\rm c}\rightarrow \infty$, we can obtain from (\ref{gammaepsilon}), (\ref{gammab}) and (\ref{gammasigma}) 
\begin{eqnarray}
& &1+\gamma\gamma^\dagger=1+\bar\gamma\bar\gamma^\dagger\nonumber\\
&=&\frac{4C {\rm e}^{u(s)}}{\iu v}\left\{\epsilon-E_{\rm p}(b)-\frac{1}{C}\int^\infty_0\rmd s \tilde\Sigma(s){\rm e}^{-u(s)}\right\}.
\label{denominator}
\end{eqnarray} 
In (\ref{denominator}), the second term in the curly bracket has been given by (\ref{Epb}).  

Now we calculate the third term in $\left\{\cdots \right\}$ in (\ref{denominator}). 
First we note that the off-diagonal elements $\bar\Sigma_{12}+\bar\Sigma_{21}$ 
vanishes on the trajectory with $b=0$ and $\tilde\Sigma$ can be replaced by $\Sigma_{11}$. 
With the use of (\ref{Sigma2}) and (\ref{average1}), we obtain
\begin{eqnarray}
& &\frac{1}{C}\int^\infty_0\rmd s \Sigma_{11}(s){\rm e}^{-u(s)}\nonumber\\
&\sim&\frac{\pi v\Gamma_{\rm n}}{2\iu C^2}\int_0^\infty\rmd s\frac{{\rm e}^{-2u(s)}}{\sqrt{E'^2s^2-\tilde\epsilon^2}}.  
\label{integral1}
\end{eqnarray}
For $s\sim \xi_0$, $u(s)\sim 2\Delta_\infty s/v$. Hence, the above expression can be reduced to
\begin{eqnarray}
& &\frac{\pi v\Gamma_{\rm n}}{2\iu C^2}\int_0^{v/(4\Delta_\infty)}\rmd s\left(E'^2s^2-\tilde\epsilon^2\right)^{-1/2}\nonumber\\
&=&\tilde\Gamma_{\rm n}\left[\frac1\iu\ln\left(\frac{\tilde\Delta_\infty}{2\left|\tilde\epsilon\right|}\right)+\cot^{-1}\frac{{\rm Im}\tilde\epsilon}{{\rm Re}\tilde\epsilon}+{\cal O}\left(\frac{\tilde\epsilon^2}{\tilde\Delta_\infty^2}\right)\right], 
\label{integral2}
\end{eqnarray}
with $\tilde\Gamma_{\rm n}=\pi v\Gamma_{\rm  n}/(2C^2E')$. From ({\ref{denominator}), ({\ref{integral1}) and ({\ref{integral2}), we immediately see that $1+\gamma\gamma^\dagger$ vanishes at
\begin{equation}
{\rm Re}\epsilon=E_{\rm p}+\pi\tilde\Gamma_{\rm n}/2,\\
{\rm Im}\epsilon=-\tilde\Gamma_{\rm n}\ln\left(\frac{\tilde\Delta_\infty}
{2E_{\rm p}}\right)\equiv -\Gamma. 
\label{pole1}
\end{equation}
Now we examine the consistency of our results. The assumption (\ref{assumption1}) is obviously satisfied in (\ref{pole1}).

The quantity $\tilde \Gamma_{\rm n}$ is of the order of $\Gamma_{\rm n}/\ln\left(\pi\xi_0/(2\xi_1)\right)$. Therefore, the condition $\Gamma\ll \Delta_\infty$ gives
\begin{eqnarray}
\Gamma/\Delta_\infty&\sim&\left(\tilde\Gamma_{\rm n}/\Delta_\infty\right)\ln\left(\Delta_\infty/E_{\rm p}\right)\nonumber\\
&\sim&\left(\Gamma_{\rm n}/\Delta_\infty\right)\ln\left(
b/\xi_0\right)/\ln\left(\pi\xi_0/(2\xi_1)\right)\ll 1.  
\label{additional1}
\end{eqnarray}
Our assumption (\ref{assumption2}) on the self-energy can be rewritten as
\begin{equation}
\Gamma_{\rm n}/{\rm Max}\left(E_{\rm p}, |\epsilon|\right)\ll 1. 
\label{additional2}
\end{equation}
 The condition (\ref{Gammaassumption}) is thus confirmed by
$$
\frac{1}{E'}\frac{\rmd \Gamma }{\rmd b}=\frac{\tilde \Gamma_{\rm n}}{E' b}
\sim \frac{\Gamma_{\rm n}}{E_{\rm p}\ln\left(\pi\xi_0/(2\xi_1)\right)}\ll 1. 
$$ 

Now we can write down the final expression for $\hat g$. 

The matrix part of $\hat g$ in (\ref{twobytwo}) has a non-vanishing value for $\epsilon=b=\Sigma=0$ 
\begin{equation}
\left(
\begin{array}{cc}
1-\gamma\gamma^\dagger&2\gamma\\
2\gamma^\dagger&-1+\gamma\gamma^\dagger
\end{array}\right)\rightarrow
2\hat M_0\left(\alpha\right).
\label{matrixlimit}
\end{equation}

As a result, we obtain 
\begin{equation}
\hat g\sim \frac{\pi v {\rm e}^{-u(s)}\hat M_0\left(\alpha\right)}{2C\left(\epsilon-E_{\rm p}-\pi\tilde\Gamma_{\rm n}/2+\iu\Gamma\right)}
\label{ghatlow}
\end{equation}
under the conditions $(\epsilon, \Gamma_{\rm n})\ll \Delta_\infty$, $b\ll \xi_0$ and two additional ones (\ref{additional1}) and (\ref{additional2}).   
\section{P-wave Case}
In the present section, we consider p-wave superconductors with the pair potential 
\begin{equation}
\Delta({\mib r})=\Delta_0(r){\rm e}^{\iu (\phi\pm \alpha)}
\label{Deltap}
\end{equation}
 in a way similar to the previous case.
\subsection{$(p_x+\iu p_y)$ superconductors without impurities}
First we consider the green function in the pure case ($\hat \Sigma=0$). The spectrum within a core of pure superconductors with ${\mib d}=\hat{\mib z}(p_x\pm \iu p_y)$ has been calculated in ref.~\citen{Matsumoto} numerically using the Bogoliubov-de Gennes equation. The result in the present subsection {\it 4.1} corresponds to the $k_{\rm F}\xi_0 \gg 1$ limit of that in ref.~\citen{Matsumoto}. 

If we use the same parameterization (\ref{twobytwo}) for $\hat g$ and the expression (\ref{Deltap}) and set $\Sigma=0$, equations (\ref{Riccati1}a,b) still hold. Further when we introduce the following notations:
\begin{equation}
\begin{array}{ccc}
\gamma=\bar\gamma {\rm e}^{\iu 2\alpha},&\gamma^\dagger=\bar\gamma^\dagger {\rm e}^{-\iu 2\alpha},&\Delta=\bar\Delta {\rm e}^{\iu 2\alpha},\\
\Delta^*=\bar\Delta^\dagger {\rm e}^{-\iu 2\alpha},  
\end{array}
\label{bargammap}
\end{equation}
the resulting $\bar\gamma^{(\mbox{ },\dagger)}$ in (\ref{bargammap}), then, turn out be exactly same as those in the previous section. We can thus use (\ref{gammaepsilon}) and (\ref{gammab}) to obtain the solutions for small $\epsilon$ and small $b$. As a result, the expression for the quasiclassical Green of pure p-wave superconductors is given by
\begin{equation}
\hat g\sim \frac{\pi v {\rm e}^{-u(s)}\hat M_+}{2C\left(\epsilon-E_{\rm p}+\iu\delta\right)}, 
\label{ghatlowp}
\end{equation}
with 
\begin{equation}
\hat M_+=\hat M_+\left(\alpha\right)\equiv 
\left(
\begin{array}{cc}
1&-\iu {\rm e}^{\iu 2\alpha}\\
-\iu {\rm e}^{-\iu 2\alpha}&-1. 
\end{array}\right)
\label{Mplus}
\end{equation}
The fact that the phase factor in the off-diagonal part in $\hat g$ is not $\pm \alpha$ but $\pm 2\alpha$ leads to a new aspect in the present case. 
\subsection{$(p_x+\iu p_y)$ superconductors in the Born limit}
Now we consider the case with the self-energy. We assume again that $\hat g$ has approximate form of 
\begin{equation}
\hat g\sim \frac{\pi v {\rm e}^{-u(s)}\hat M_+}{2C\left(\epsilon-E_0-E'b+\iu\Gamma\right)}. 
\label{ghatlowp2}
\end{equation}
Following the same procedure as the previous section, we obtain from (\ref{ghatlowp2}) the expressions for self-energy
\begin{eqnarray}
& &{\rm e}^{-\iu 2\phi}\Sigma_{12}={\rm e}^{\iu 2\phi}\Sigma_{21}\nonumber\\
&\sim& -\frac{\pi v {\rm e}^{-u(\tilde s_1)}}{2C E'r}\left(\frac{1-2w^2}{\sqrt{1-w^2}}+2\iu w\right). 
\label{selfenergyp}
\end{eqnarray}         
For the diagonal part $(\Sigma_{11}, \Sigma_{22})$, the expressions are the same as those for s-wave. 

Now we solve the quasiclassical equation with use of parameterization (\ref{twobytwo}) ;
\begin{equation}
\hat g=\frac{-\iu\pi}{1+\bar\gamma\bar\gamma^\dagger}
\left(
\begin{array}{cc}
1-\bar\gamma\bar\gamma^\dagger&2\bar\gamma {\rm e}^{\iu2\alpha}\\
2\bar\gamma^\dagger {\rm e}^{-\iu 2\alpha}&-1+\bar\gamma\bar\gamma^\dagger
\end{array}\right).
\label{twobytwop2}
\end{equation}
$\bar\gamma$ and $\bar\gamma^\dagger$ in (\ref{twobytwop2}) are the solutions of the following Riccati differential equations: 
\begin{subeqnarray}
&\iu {\mib v}\cdot{\mib \nabla}\bar\gamma&=-\left(\bar\Delta^\dagger-\bar\Sigma_{21}\right)\bar\gamma^2-2\left(\epsilon-\Sigma_{11}\right)\bar\gamma-\left(\bar\Delta+\bar\Sigma_{12}\right)\label{Riccatip1}\\
&\iu {\mib v}\cdot{\mib \nabla}\bar\gamma^\dagger&=-\left(\bar\Delta+\bar\Sigma_{12}\right)\bar\gamma^{\dagger 2}+2\left(\epsilon-\Sigma_{11}\right)\bar\gamma^\dagger-\left(\bar\Delta^\dagger-\bar\Sigma_{21}\right),  
\label{Riccatip2}
\end{subeqnarray}
where $\bar\Delta$ and $\bar\Delta^\dagger$ have been defined in (\ref{bargammap}). $\bar\Sigma_{12}$ and $\bar\Sigma_{21}$ are defined as
\begin{equation}
\Sigma_{12}=\bar\Sigma_{12}{\rm e}^{\iu 2\alpha},\quad \Sigma_{21}=\bar\Sigma_{21}{\rm e}^{-\iu 2\alpha}. 
\end{equation}
Equations (\ref{Riccatip1}a,b) should be solved under the initial conditions:
\begin{subeqnarray}
\bar\gamma(s=-s_{\rm c})&=&-\bar\Delta(s=-s_{\rm c})/\left(\bar\epsilon+\iu\sqrt{\left|\Delta\right|^2-\bar\epsilon^2}\right),\label{icp1a}\\
\bar\gamma^\dagger(s=s_{\rm c})&=&\bar\Delta^\dagger(s=s_{\rm c})/\left(\bar\epsilon+\iu\sqrt{\left|\Delta\right|^2-\bar\epsilon^2}\right), 
\label{icp1b}
\end{subeqnarray}
where $\bar\epsilon$ is given as the solution of   
\begin{equation}
\bar\epsilon\left(1-\pi\Gamma_{\rm n}/\sqrt{\left|\Delta\right|^2-\bar\epsilon^2}\right)=\epsilon. 
\end{equation}  
The initial conditions (\ref{icp1a}a,b) are derived from the bulk solution for p-wave superconductors. Nonmagnetic impurities give pair-breaking effects on the non-swave homogeneous superconductors. The presence of impurities, therefore, affects the bulk solution through $\bar\epsilon$.  

What we need are the solutions of (\ref{Riccatip1}a,b) for small $\epsilon$, $b$ and $\Gamma_{\rm n}$. The initial conditions (\ref{icp1a}a,b) are expanded as
\begin{eqnarray}
\bar\gamma(s=-s_{\rm c})&\sim&\bar\gamma^\dagger(s=s_{\rm c})\nonumber\\
&\sim&-\iu+\frac{\epsilon}{|\Delta_\infty|}-\frac{b}{s_{\rm c}}+\frac{{\cal O}\left(\epsilon\Gamma_{\rm n}, \epsilon^2\right)}{|\Delta_\infty|^2}.
\label{icp2}
\end{eqnarray}
The expression (\ref{icp2}) shows that initial conditions for impure superconductors are the same as those for pure superconductors in the case of small $\epsilon$, $b$ and $\Gamma_{\rm n}$; this fact suggests that impurity effects should be different from those for bulk superconductors.  

Now we calculate $\bar\gamma$ and $\bar\gamma^\dagger$ for p-wave superconductors. All the results from (\ref{expand}) to (\ref{denominator}) apply to the present case and hence we can begin with the expression (\ref{denominator}) for $1+\bar\gamma\bar\gamma^\dagger$.   

From (\ref{selfenergyp}) for $\Sigma_{12}$ and $\Sigma_{21}$, it follows that $\bar\Sigma_{12}=\bar\Sigma_{21}$ on the trajectory with $b=0$. As a result, those off-diagonal elements {\it contribute} to the bound state pole in p-wave case, in contrast to the s-wave case. 

The integral
\begin{equation}
\frac{-1}{C}\int_0^\infty\rmd s\tilde\Sigma(s){\rm e}^{-u(s)}
\end{equation}   
for p-wave case can be reduced to 
\begin{equation}
\frac{\pi v\Gamma_{\rm n}\tilde\epsilon}{C}\int_0^\infty\rmd s\frac{{\rm e}^{-2u(s)}}{\sqrt{E'^2s^2-\tilde\epsilon^2}\left(\sqrt{E'^2s^2-\tilde\epsilon^2}-\iu \tilde \epsilon\right)}.
\label{integral3}
\end{equation}
The denominator of the integrand in (\ref{integral3}) makes the integrand rapidly convergent. Therefore we can replace ${\rm e}^{-2u(s)}$ by unity. After performing integration, (\ref{integral3}) becomes $\iu 2\tilde\Gamma_{\rm n}$. Consequently, we obtain 
\begin{equation}
E_0=0,\quad E'=E_{\rm p}/b,\quad \Gamma=2\tilde\Gamma_{\rm n}
\label{pwaveresult}
\end{equation}
for the green function (\ref{ghatlowp2}) of p-wave superconductors. 

If we compare (\ref{pwaveresult}) with (\ref{pole1}), we can easily see a qualitative difference of $\Gamma$ between the two cases; $\Gamma$ in s-wave case has logarithmic dependence on the bound state energy while that in p-wave case is energy-independent. This difference comes from different behaviour of $\tilde \Sigma$ in (\ref{denominator}) and (\ref{integral3}). In s-wave case, $\tilde\Sigma(s)=\Sigma_{\rm s}(s)\sim 1/s$ for $\left|\tilde\epsilon\right|/E'\ll s\ll\xi_0$. This long tail gives the logarithmic dependence $\ln(\xi_0 E'/\left|\tilde\epsilon\right|)$ in $\Gamma$. In $\tilde \Sigma(s)$ of p-wave superconductors, on the other hand, the long-tail of $\Sigma_{11}(s)$ is cancelled by that of $\Sigma_{12}$. This leads to energy-independent $\Gamma$.
\subsection{$(p_x+\iu p_y)$ superconductors away from the Born limit}
In this subsection, we consider the impurity scattering rate within the t-matrix approximation (\ref{Sigmatmatrix}); the phase shift $\delta_0$ is to be small $(|\delta_0|\ll 1)$ but finite. Following the same method as that in the previous one, we obtain the green function in the form of (\ref{ghatlowp2}) with 
$E_0=0$, $E'=E_{\rm p}/b$ and 
\begin{equation}
\frac{\Gamma}{\tilde\Gamma_{\rm n}}\sim 
\left\{
\begin{array}{cr}
2&\mbox{ for } E_\delta \ll E_{\rm p}\ll \Delta_\infty\\
\left(2\pi/3^{3/2}\right)\left(E_{\rm p}/E_\delta\right)^{2/3}&\mbox{ for } \tilde \Gamma_{\rm n}^3/E_\delta^2\ll E_{\rm p}\ll E_\delta, 
\end{array}
\right. 
\label{Gammatmatrix}
\end{equation}
with $E_\delta=v\left|\delta_0\right|/C \sim \left|\delta_0\right| \Delta_\infty$. We will describe the derivation of (\ref{Gammatmatrix}) in Appendix. 

We see that finite phase shift sets an energy scale $E_\delta$, above which the Born approximation works well and below which it breaks down. 
The suppression of the impurity scattering rate in the energy region $E_{\rm p}\ll E_\delta$ may be related to ^^ ^^ the spectrum rigidity" within a vortex core chiral superconductors against a single impurity~\cite{Volovik}; The approximation in ref.~\citen{Volovik}, where only the backward scattering is taken into account, seems to be valid in that energy region. 

On the other hand, in the energy region $E_\delta \ll E_{\rm p}\ll \Delta_\infty$, the impurity scattering is large; this suggests the strong level mixing and hence we expect that the random-matrix approach~\cite{Bocquet,Ivanov} should be applicable in this energy region.          
\subsection{$(p_x-\iu p_y)$ superconductors}
We consider in this subsection the impurity effects in $(p_x-\iu p_y)$ superconductors under the test pair-potential
\begin{equation}
\Delta({\mib r})=\Delta_0(r){\rm e}^{\iu\left(\phi-\alpha\right)}. 
\label{testpotentialminus}
\end{equation}
We will see that the impurity effect in vortex core of superconductors with time-reversal-symmetry-breaking is {\it chirality-sensitive}.

For those superconductors without impurity, we obtain in the same method for the green function
\begin{equation}
\hat g\sim \frac{\pi v {\rm e}^{-u(s)}\hat M_-}{2C\left(\epsilon-E_{\rm p}+\iu\delta\right)}, 
\label{ghatlowpminus}
\end{equation}
with 
\begin{equation}
\hat M_-\equiv \left(
\begin{array}{cc}
1&-\iu \\
-\iu &-1. 
\end{array}\right)
\label{Mminus}
\end{equation}
We note that the off-diagonal elements of the matrix part in (\ref{ghatlowpminus}) have no $\alpha$-dependence. This fact will turn out to be crucial. 

On the basis of (\ref{ghatlowpminus}), we expect the green function of the form
\begin{equation}
\hat g\sim \frac{\pi v {\rm e}^{-u(s)}\hat M_-}{2C\left(\epsilon-E_0-E'b+\iu\Gamma\right)}
\label{ghatlowpminus2}
\end{equation}
in the presence of impurities. 
From (\ref{ghatlowpminus2}), we immediately see that 
\begin{equation}
\Sigma_{12}=\Sigma_{21}=-\iu\Sigma_{11}.
\label{relation}
\end{equation}    
By following the argument from (\ref{expand}) to (\ref{denominator})
with replacement of $(\bar\gamma^{(\mbox{ },\dagger)},\bar\Sigma)$ by $(\gamma^{(\mbox{ },\dagger)},\Sigma)$
, we obtain
for $E_0$ and $\Gamma$ in (\ref{ghatlowpminus2})
\begin{equation}
E_0=0,\quad \Gamma=\frac{\iu}{C}\int_0^\infty\rmd s \tilde\Sigma(s){\rm e}^{-u(s)}. 
\label{Gammapminusequation}
\end{equation} 
The relation (\ref{relation}), however, yields $\tilde\Sigma=0$ and accordingly, $\Gamma$ in (\ref{Gammapminusequation}) vanishes. Our calculation is based on the smallness of $|\epsilon|/\Delta_\infty$ and $b/\xi_0$. Therefore we expect that at most
\begin{equation}
\Gamma\sim {\cal O}\left(\frac{\Gamma_{\rm n}E_{\rm p}}{\Delta_\infty}\right)
\label{Gammapminus}
\end{equation}
for $(p_x-\iu p_y)$ superconductors with the test potential (\ref{testpotentialminus}). 

The result (\ref{Gammapminus}) is unchanged even away from the Born limit within the t-matrix approximation. Crucial is the structure of $\hat M_-$ in $\hat \Sigma$; it is common for Born and t-matrix approximations. 

A comparison between (\ref{pole1}) and (\ref{Gammapminus}) clearly shows that the impurity effect in vortex of superconductors with time-reversal-symmetry-breaking depends severely on the chirality for $E_\delta\ll E_{\rm p}\ll \Delta_\infty$ even under the same trial amplitude $\Delta_0(r)$ and vorticity of the pair-potential. 

In the vortex core with $+1$ vorticity of $(p_x-\iu p_y)$ superconductors, the result in ref.~\citen{Volovik} seems to be valid in wider energy region while the random-matrix approach becomes less applicable,  compared to the $(p_x+\iu p_y)$ case with +1 vorticity.  
\section{Implication for Flux Flow Ohmic Conductivity}
Now we consider the implications of our results to the flux flow conductivity. The temperature dependence of the flux flow conductivity in moderately clean regime (\ref{moderatelyclean}) is governed mainly by $T$-dependences of $\Gamma(E_{\rm p}=T)$ and $\Delta_0(r)$ (or $\xi_1$ in (\ref{xi1})). The ratio $\xi_1/\xi_0$ is expected to be~\cite{KramerPesch}  $\sim T/\Delta_\infty$ (Kramer-Pesch effect) if the zero energy Andreev bound states exist on the quasiclassical trajectory with vanishing impact parameter~\cite{Volovik2}. Here we assume that the Kramer-Pesch effect occurs. Under this assumption, the expression (\ref{sigmaf2}) reduces to 
\begin{eqnarray}
\sigma_{\rm f}/\sigma_{\rm n}&\sim& \ln\left(\frac{\Delta_\infty}{T}\right)\frac{H_{\rm c2}\Gamma_{\rm n}}{B\Gamma(E_{\rm p}=T)}\nonumber\\
&\sim&\left[\ln\left(\frac{\Delta_\infty}{T}\right)\right]^2\frac{H_{\rm c2}\tilde \Gamma_{\rm n}}{B\Gamma(E_{\rm p}=T)}.
\label{sigmaf3}
\end{eqnarray}    
For the Born limit of s-wave case, the substitution of (\ref{pole1}) into (\ref{sigmaf3}) reproduces the known result (\ref{cleans})~\cite{LO76,KopninLopatin}. For p-wave with ${\mib d}=\hat{\mib z}\left(p_x+\iu p_y\right)$, we obtain from (\ref{sigmaf3}) and (\ref{Gammatmatrix})
\begin{eqnarray}
\sigma_{\rm f}/\sigma_{\rm n}
&\sim& \left[\ln\left(\Delta_\infty/T\right)\right]^2\left(H_{\rm c2}/B\right)\nonumber\\
& &\left\{
\begin{array}{cr}
{\cal O}(1)&\mbox{ for } E_\delta \ll T\ll \Delta_\infty\\
\left(E_\delta/T\right)^{2/3}&\mbox{ for } \Gamma_{\rm n}^3/E_\delta^2\ll T\ll E_\delta.  
\end{array}
\right. 
\label{sigmaf4}
\end{eqnarray} 
Thus we expect novel enhancement $\sigma_{\rm f}$ at temperatures lower than $E_{\delta}$ for the present case. 

For p-wave with ${\mib d}=\hat{\mib z}\left(p_x-\iu p_y\right)$, on the other hand, the impurity scattering rate $\Gamma$ is much smaller than $\Gamma_{\rm n}$ in energy region $E\ll\Delta_\infty$. The flux flow conductivity $\sigma_{\rm f}$ is, therefore, expected to be much larger than $\sigma_{\rm n}\left[\ln\left(\Delta_\infty/T\right)\right]^2\left(H_{\rm c2}/B\right)$ at temperatures much lower than $\Delta_\infty$.  
\section{Discussion}
So far we have seen that the impurity scattering rate of the Andreev bound states in vortex core depends sensitively on the parity and chirality of pair-potential. We discuss this point in more intuitive way in the present section. 

The impurity scattering rate of an in-going state is, in general, detemined by two factors: the density of states of available out-going states and the matrix element. We considered s- and p- wave superconductors having common pair-potential amplitude ($\Delta_0(r)$) and spectrum of the Andreev bound states. Therefore, the phase-sensitive scattering rate should originate from the matrix element. In the course of calculations, actually, we note that different structures of matrix parts (\ref{M0}) (\ref{Mplus}) (\ref{Mminus}) of the green function are crucial factors giving different scattering rates. Difference among the three matrices lies in the off-diagonal element, which is nothing but the ^^ ^^ Riccati amplitude " $\gamma$ or $\gamma^\dagger$ of zero energy bound states on the trajectory at zero impact parameter in pure superconductors. It has been mentioned~\cite{Schopohl,Eschrig2} that the $\gamma$ and $\gamma^\dagger$ in pure superconductors correspond to the ratio of particle amplitude $u$ to the hole one $v$ of the Andreev equation~\cite{Andreev}. In low energy limit, the amplitude of the two are the same and the relative phase is fixed strictly by the phase of the pair-potential so that the multiple Andreev reflections within the core constructively interfere. The phase of the pair-potential along the trajectory with zero impact parameter, of course, depends on the parity of pairing symmetry.       
Let us consider the scattering rate due to Born scatterers of a bound state with energy $\epsilon$ with impact parameter $b_\epsilon\equiv \epsilon/E'$. The scattering rate of a bound state is given by the integration of the scattering rate at each point along the trajectory. 
The density of states of available out-going state is given by the local density of states $\langle (\epsilon-E_{\rm p}+\iu \delta)^{-1}\rangle$, which is proportional to $1/s$ for $b_\epsilon\ll s\ll \xi_0$. This factor alone gives $\Gamma/\Gamma_{\rm n}\sim \ln(\xi_0 E'/\epsilon)$. This expectation reproduces the result in s-wave case~(\ref{pole1}) and do not give the results in p-wave cases. Owing to the long tail of the local density of states as a function of $s$, the main contribution to the scattering rate comes from the large-$s$ region, where the scattering angle is near $\pi$; backward scattering. From the argument of the previous paragraph, it is understandable that the strength of the backward scattering between the Andreev bound states in vortex core is parity-dependent. 

Chirality dependence of the scattering rates in the Born limit comes from the small $s$ region (forward scattering region). The phase of the pair-potential has contributions from vorticity and chirality. Owing to the circular symmetry in the present study, the completely destructive interference occurs in the matrix element in the case of $\Delta({\mib r})=\Delta_0 {\rm e}^{\iu\left(\phi-\alpha\right)}$. 
 
Next we consider the non-Born case of $\Delta({\mib r})=\Delta_0 {\rm e}^{\iu\left(\phi+\alpha\right)}$. The factor $\langle \tilde f \rangle\langle f \rangle-\langle g\rangle^2$ in the denominator of (\ref{Sigmatmatrix}) behaves as $(\xi_0/s)^3$. Therefore the main effect of $|\delta_0|$ is the suppression of the scattering rate in the region $b_\epsilon\le s\ll s_0\equiv |\delta_0|^{2/3}\xi_0$. This fact explains why the scattering rate decreases appreciably and chirality dependence becomes small in the energy region lower than $E_\delta$.  

Lastly, we make a remark on the formulation used in \S~3 and \S~4. The impurity scattering rates of the Andreev bound states within s-wave vortex core was already calculated in ref.~\citen{Kopnin99} in a different method. Application of the method used in ref.~\citen{Kopnin99} to p-wave case reproduces the result in \S 4. In this sense, the two methods are equivalent. The meaning of the approximation is, however, clearer in the present work than in the earlier work. 
 
The method in ref.~\citen{Kopnin99} was a generalization of that of Kramer and Pesch~\cite{KramerPesch}, which is originally developed for pure superconductors. The spirit of the crucial assumption in the original Kramer-Pesch paper (which is written in the sentense immediately below eq.~(14) in ref.~\citen{KramerPesch}) is not obvious as it stands. The method developed in the present paper is, on the other hand, a generalization of that presented in Appendices G and H of ref.~\citen{Eschrig} to impure superconductors. While the Green function becomes singular near the bound state pole, Riccati amplitudes $\gamma$, $\gamma^\dagger$ are regular as functions of energy $\epsilon$ and impact parameter $b$. That is why  we can expand those amplitudes safely with $\epsilon$ and $b$ and the spirit and validity of approximation are clear. This ^^ ^^ good" behaviour of Riccati amplitude comes from the behaviour of wave functions; we recall that the Riccati amplitudes are related directly to the wave functions of Andreev equation in pure case. Of course, wave function cannot be defined any more after the impurity-average. The Reccati amplitudes are, however, well-defined quantities analogous to wavefunctions even in the presence of energy dissipation. Riccati formalism of quasiclassical theory of superconductivity turns out to be powerful in analytical as well as numerical study.

\section{Conclusion}
In summary, we presented the calculations of the impurity scattering rate of Andreev bound states within 2D pancake vortex of s-wave and chiral p-wave moderately clean superconductors and further considered flux flow conductivity.

We found that energy dependence of impurity scattering rate is governed by the relative phase between the particle and hole channels in the wavefunction of an Andreev bound state. This relative phase is strictly fixed by the parity and chirality of pair-potential and thus the scattering rate and resulting flux flow conductivity become parity- and chirality-sensitive in the moderately clean unconventional superconductors. The calculation of the impurity scattering rate under self-consistently determined pair-potential and self-energy is left as a future problem.   
\acknowledgement
I acknowledge D. Rainer for his instructive discussions on the quasiclassical theory of superconductivity and bringing my attention to refs.~\citen{Nagato1,Higashitani,Nagato2,Schopohl,Eschrig,Eschrig2}. 
This work is supported by Grant-in-Aid for Scientific Research on Priority Areas (A) of ^^ ^^ Novel Quantum Phenomena in Transition Metal Oxides" (12046225) from the Ministry of Education, Science, Sports and Culture.    
\appendix
\section{Derivation of (\ref{Gammatmatrix})}
In this appendix, we calculate the impurity scattering rate of Andreev bound states in vortex core (with +1 vorticity) of p-wave superconductors with ${\mib d}=\hat{\mib z}\left(p_x+\iu p_y\right)$ within the t-matrix approximation. We consider the case where $|\delta_0|\ll 1$; we hence replace $\cos\delta_0$ by unity and $\sin\delta_0$ by $\delta_0$ in the following part.  
  
Following the same proceduce as that used in \S 4.2, we obtain 
\begin{equation}
\hat g\sim \pi v {\rm e}^{-u(s)}\hat M_+\left[2C\left(\epsilon -E_{\rm p}+S\right)\right]^{-1}
\end{equation}
with 
\begin{equation}
S=-\frac{1}{C}\int_0^\infty \rmd s\tilde \Sigma(s){\rm e}^{-u(s)}. 
\label{S}
\end{equation}
The imaginary part of $S$ yields the impurity scattering rate $\Gamma$. 
Further the expression (\ref{S}) reduces to
$$
\tilde\Gamma_{\rm n}I\left(E_\delta/\tilde \epsilon\right). 
$$
Here $I(\lambda)$ is defined, for complex $\lambda$ satisfying ${\rm Re}\lambda>0$ and ${\rm Im}\lambda<0$, as  
\begin{equation}
I(\lambda)
=-\int_{-\infty \lambda}^{+\infty \lambda}\rmd z \frac{\zeta(z)+1}{\zeta(z)\left(\zeta(z)+1\right)^2-\lambda^2},
\label{I}
\end{equation}
where $\zeta(z)$ is given by
\begin{eqnarray}
& &\iu \zeta(z)=\sqrt{z^2-1}\nonumber\\
&=&\left|z^2 -1\right|^{1/2}\exp\left[\iu \left\{{\rm Arg}(z-1)+{\rm Arg}(z+1)\right\}/2\right].
\label{zetadefinition} 
\end{eqnarray}
Principal values of arguments are, respectively, taken as 
\begin{equation}
0<{\rm Arg}(z-1)\le 2\pi,\quad -\pi\le{\rm Arg}(z+1)<\pi.
\label{Arg}
\end{equation}
 Corresponding branch cuts are given by ${\rm Im}z=0$ and $|{\rm Re}z|>1$. The path of integral in (\ref{I}) is taken within the Riemann surface specified by (\ref{Arg}). From now on, we impose the condition that $-\pi/4<{\rm arg}\lambda\le 0$, which is equivalent to $0\le {\rm arg}\tilde \epsilon<\pi/4$ and evaluate the integral in the two limits $|\lambda|\gg 1$ and $|\lambda|\ll 1$. In both limits, the integral (\ref{I}) can be written as the sum of the residue of the pole of the integrand and the integral along a branch cut;
$$
I(\lambda)=I_{\rm pole}(\lambda)+I_{\rm cont}(\lambda).
$$
The contribution from the pole $z_0$ which yields 
\begin{equation}
\zeta(z_0)\left(\zeta(z_0)+1\right)^2=\lambda^2
\label{pole}
\end{equation}
 turns out to be  
$$
I_{\rm pole}(\lambda)=\frac{2\pi \iu \zeta(z_0)}{z_0 \left(3\zeta(z_0) +1\right)}.
$$
The other contribution $I_{\rm cont}$ is given by 
\begin{equation}
I_{\rm cont}(\lambda)=-2\iu \int_0^\infty\rmd y\frac{y^2\left(y^2+1+\lambda^2\right)}{\sqrt{y^2+1}\left\{\left(y^3-y\right)^2+\left(2y^2 +\lambda^2\right)^2\right\}}. 
\label{Icont}
\end{equation}

First we consider the asymtotics of $I_{\rm pole}(\lambda)$. 
In the limit $|\lambda|\ll 1$, it follows from (\ref{zetadefinition}) and (\ref{pole}) that $\zeta(z_0)\rightarrow \lambda^2$ and $z_0\rightarrow 1$. We immediately see that 
\begin{equation}
I_{\rm pole}(\lambda)\rightarrow 2\pi \iu \lambda^{2} \mbox{ for }  |\lambda|\rightarrow 0.
\label{Ipolelimit1}
\end{equation}
When $|\lambda|\gg 1$, on the other hand, we obtain that $\zeta(z_0)\rightarrow \lambda^{2/3}$ and $z_0\rightarrow \iu \lambda^{2/3}$. Here the branch cut in $\lambda$ plane is taken along the real axis with ${\rm Re}\lambda<0$. These results lead to 
\begin{equation}
I_{\rm pole}(\lambda)\rightarrow 2\pi\lambda^{-2/3}/3, \mbox{ for }|\lambda|\rightarrow \infty.
\label{Ipolelimit2}
\end{equation}

Next consider the behavior of $I_{\rm cont}(\lambda)$ in the two limits. 
Around $\lambda=0$, $I_{\rm cont}(\lambda)$ is a regular function of $\lambda$ and is expanded as
\begin{equation}
I_{\rm cont}(\lambda)=-2\iu +\frac{44\iu \lambda^2}{15}+{\cal O}\left(\lambda^4\right). 
\label{Icontlimit1}
\end{equation} 
When $|\lambda|\gg 1$, on the other hand, the main contribution in (\ref{Icont}) comes from the region $y\gg 1$. We can, therefore, ignore unity in the integrand and (\ref{Icont}) reduces to 
\begin{equation}
-\iu \int_0^\infty \rmd \eta \frac{\rmd \eta \left(\eta+\lambda^2\right)}{\eta\left(\eta-1\right)^2 +4\eta^2 +4\lambda^2\eta+\lambda^4}. 
\label{etaintegral}
\end{equation} 
The integrand in (\ref{etaintegral}) approaches $\lambda^2/\left(\eta^3 +\lambda^4\right)$ for $\eta\ll |\lambda|^2$ and $1/(\eta +4\lambda^2)$ for $|\lambda|^2\ll \eta$. The former contribution to (\ref{etaintegral}) turns out to be
\begin{equation}
\frac{1}{\iu}\int_0^{|\lambda|^2}\frac{\lambda^2\rmd \eta}{\eta^3 +\lambda^4}\sim \frac{1}{\iu} \int_0^\infty \frac{\lambda^2\rmd \eta}{\eta^3 +\lambda^4}=-\frac{2\pi \iu \lambda^{-2/3}}{3\sqrt{3}}, 
\label{Icontlimit2}
\end{equation} 
where the branch cut in $\lambda$ plane is taken along the real axis with ${\rm Re}\lambda<0$. On the other land, the contribution from the region  $|\lambda|^2\ll \eta$ can be easily shown to be ${\cal O}(1/\lambda)$. Therefore (\ref{Icontlimit2}) gives the asymptotics of $I_{\rm cont}$ in the limit of $|\lambda|\rightarrow \infty$.  

Now we find from (\ref{Ipolelimit1}) and (\ref{Icontlimit1}) that $S$ in (\ref{S}) becomes
\begin{equation}
-2\iu \tilde \Gamma_{\rm n}\left(1+{\cal O}((\tilde \epsilon/E_\delta)^2)\right), \mbox{ for }|\tilde \epsilon |\gg E_\delta
\label{slimit1}
\end{equation}
 and from (\ref{Ipolelimit2}) and (\ref{Icontlimit2}) that 
\begin{equation}
2\pi/3\left(1-\iu /\sqrt{3}\right) \tilde \Gamma_{\rm n}\left(\tilde \epsilon/E_\delta\right)^{2/3},\mbox{ for }|\tilde \epsilon |\ll E_\delta.
\label{slimit2}
\end{equation}

Expression (\ref{slimit1}) immediately leads to $\Gamma\sim 2\tilde\Gamma_{\rm n}$; results in (\ref{Gammatmatrix}) for that energy region. In (\ref{slimit2}), on the other hand, a real part exists, which imposes additional restriction on the validity of our results. We have assumed that the dispersion of the bound state energy is proportional to the impact parameter $b$ (see (\ref{assumption1}) in \S 3). This assumption holds only if the condition 
\begin{equation}
\tilde \epsilon\gg \tilde \Gamma_{\rm n}\left(\tilde \epsilon/E_\delta\right)^{2/3}
\label{condition}
\end{equation}
is satisfied. We note that $\tilde\epsilon\sim  E_{\rm p}$ in the vicinity of the bound state pole. Consequently, we obtain (\ref{Gammatmatrix}) from (\ref{slimit2}) and (\ref{condition}).


\begin{thebibliography}{99}
\bibitem{Caroli}C.~Caroli, P. G. de Gennes and J. Matricon: Phys. Lett. {\bf 9} (1964) 307. 
\bibitem{BardeenStephen}J. Bardeen and M. J. Stephen: Phys. Rev. {\bf 140} (1965) A1197.
%
%
\bibitem{BardeenSherman}J. Bardeen and R. D. Sherman: Phys. Rev. B {\bf 12} (1975) 2634. 
\bibitem{LO76}A. Larkin and Yu. Ovchinnikov: Pis'ma Zh. Eksp. Teor. Fiz. {\bf 23} (1976) 210 [Sov. Phys. JETP Lett. {\bf 23} (1976) 187].
\bibitem{KK}N. B. Kopnin and V. E. Kravtsov: Pis'ma Zh. Eksp. Teor. Fiz. {\bf 23} (1976) 631 [Sov. Phys. JETP Lett. {\bf 23} (1976) 578].
\bibitem{KramerPesch}L. Kramer and W. Pesch: Z. Phys. {\bf 269} (1974) 59.
%
%
\bibitem{KopninLopatin}N. B. Kopnin and A. V. Lopatin: Phys. Rev. B {\bf 51} (1995) 15291.
\bibitem{KopninLopatin2}N. B. Kopnin and A. V. Lopatin: Phys. Rev. B {\bf 56} (1997) 766.
%
%
\bibitem{review}See. e. g. M. Sigrist and K. Ueda: Rev. Mod. Phys. {\bf 63} (1991) 239 for a review on unconventional superconductors.   
%
%
\bibitem{KopninVolovik}N. B. Kopnin and G. E. Volovik: Phys. Rev. Lett. {\bf 79} (1997) 1377.
\bibitem{Mahklin}Yu. G. Makhlin: Phys. Rev. B {\bf 56} (1997) 11872.
%
%
\bibitem{Volovik} G. E. Volovik: Pis'ma Zh. Eksp. Teor. Fiz. {\bf 70} (1999) 601 [JETP Lett. {\bf 70} (1999) 609]. 
\bibitem{Ivanov} D. A. Ivanov: cond-mat/9911147. 
\bibitem{Bocquet}M. Bocquet, D. Serban and M. Zirnbauer: cond-mat/9910480.
%
%
\bibitem{Kambe}S. Kambe, A. D. Huxley, P. Rodiere and J. Flouquet: Phys. Rev. Lett. {\bf 83} (1999) 1842. 
\bibitem{Lutke}N. L\"utke-Entrup, R. Blaauwgeers, B. Pla\c cais, P. Mathieu, Y. Simon, M. Krusius, S. Kambe and A. Huxley: Physica B {\bf 284-288} (1999) 527. 
%
%
\bibitem{Stone}M. Stone: Phys. Rev. B {\bf 54} (1996) 13222. 
\bibitem{Blatter}G. Blatter, V. B. Geshkenbein and N. B. Kopnin: Phys. Rev. B {\bf 59} (1999)14663. 
%
%
\bibitem{note}As a related work, there exists the transport theory of vortices in the moderately clean superfluid $^3$He, where the relaxation mechanism is not impurity scattering . See e. g. N. B. Kopnin and M. M. Salomaa: Phys. Rev. B {\bf 44} (1991) 9667; N. B. Kopnin: {\it ibid} {\bf 47} (1993) 14354.
%
%
\bibitem{Mackenzie}A. P. Mackenzie, R. K. W. Haselwimmer, A. W. Tyler, G. G. Lonzarich, Y. Mori, S. Nishizaki and Y. Maeno: Phys. Rev. Lett. {\bf 80} (1998) 161. 
\bibitem{Huxley}A. D. Huxley, H. Suderow, J. P. Brison and J. Flouquet: Phys. Lett. A {\bf 209} (1995) 365. 
%
%
\bibitem{Eilenberger}G. Eilenberger: Z. Phys. {\bf 214} (1968) 195. 
\bibitem{LO68}A. Larkin and Yu. Ovchinnikov: Zh. Eksp. Teor. Fiz. {\bf 55} (1968) 2262 [Sov. Phys. JETP {\bf 28} (1969) 1200]. 
\bibitem{Eliashberg}G. M. Eliashberg: Zh. Eksp. Teor. Fiz. {\bf 61} (1971) 1254 [Sov. Phys. JETP {\bf 34} (1972) 668].
\bibitem{SereneRainer}J. W. Serene and D. Rainer: Phys. Reports {\bf 101} (1983) 221. 
%
%
\bibitem{Kopnin99}N. B. Kopnin: Phys. Rev. B {\bf 60} (1999) 581.
%
%
\bibitem{Nagato1}Y. Nagato, K. Nagai and J. Hara: J. Low Temp. Phys. {\bf 93} (1993) 33.
\bibitem{Higashitani}S. Higashitani and K. Nagai: J. Phys. Soc. Jpn. {\bf 64} (1995) 549. 
\bibitem{Nagato2}Y. Nagato, S. Higashitani, K. Yamada and K. Nagai: J. Low Temp. Phys. {\bf 103} (1996) 1. 
\bibitem{SchopohlMaki} N. Schopohl and K. Maki: Phys. Rev. B {\bf 52} (1995) 490. 
\bibitem{Schopohl}N. Schopohl: in {\it Quasiclassical Methods in Superconductivity \& Superfluidity } edited by D. Rainer and J. Sauls, (1996) 88; K. Nagai: {\it ibid} (1996) 198.
\bibitem{Eschrig}M. Eschrig: Ph. D Thesis, University of Bayreuth (1997). 
\bibitem{Eschrig2}M. Eschrig: Phys. Rev. B {\bf 61} (2000) 9061. 
%
%
\bibitem{AG}A. A. Abrikosov and L. P. Gorkov: Zh. Eksp. Teor. Fiz. {\bf 39} (1960) 1781 [Sov. Phys. JETP {\bf 12} (1961) 1243].    
%
%
\bibitem{Matsumoto}M. Matsumoto and M. Sigrist: J. Phys. Soc. Jpn. {\bf 68} (1999) 724. 
%
%
\bibitem{Volovik2}G. E. Volovik: Pis'ma Zh. Eksp. Teor. Fiz. {\bf 58} (1993) 444 [JETP Lett. {\bf 58} (1993) 460].
%
%
\bibitem{Andreev}A. F. Andreev: Zh. Eksp. Teor. Fiz. {\bf 46} (1964) 1823 [Sov. Phys. JETP {\bf 19} (1964) 1228].    
%
%


\end{thebibliography}
\end{document}